\documentclass[apj]{emulateapj}
\usepackage{comment}
\usepackage{lscape}

\shorttitle{ANOMALOUS DUST EMISSION IN NGC~6946}
\shortauthors{MURPHY ET AL.}
\slugcomment{Draft version 2.1 December 14, 2009}
\journalinfo{Accepted to \apjl~Letters on December 14, 2009}

\begin{document}

\title{The Detection of Anomalous Dust Emission in the Nearby Galaxy NGC~6946}

\author{E.J.~Murphy,\altaffilmark{1} G.~Helou,\altaffilmark{2} J.J.~Condon,\altaffilmark{3} E.~Schinnerer,\altaffilmark{4} J.L.~Turner,\altaffilmark{5} R.~Beck,\altaffilmark{6} 
B.S.~Mason,\altaffilmark{3} R.-R.~Chary, \altaffilmark{1} \& L.~Armus \altaffilmark{1}}
\altaffiltext{1}{\scriptsize {\it Spitzer Science Center,} California Institute of Technology, MC 314-6, Pasadena CA, 91125; emurphy@ipac.caltech.edu} 
\altaffiltext{2}{\scriptsize California Institute of Technology, MC 314-6, Pasadena, CA 91125}
\altaffiltext{3}{\scriptsize National Radio Astronomy Observatory, 520 Edgemont Road, Charlottesville, VA 22903, USA}
\altaffiltext{4}{\scriptsize Max Planck Institut f\"{u}r Astronomie, K\"{o}nigstuhl 17, Heidelberg D-69117, Germany}
\altaffiltext{5}{\scriptsize Department of Physics and Astronomy, UCLA, Los Angeles, CA 90095. }
\altaffiltext{6}{\scriptsize Max-Planck-Institut f\"{u}r Radioastronomie, Auf dem H\"{u}gel, Bonn, Germany}

\begin{abstract}  
We report on the Ka-band ($26-40$~GHz) emission properties for 10 star-forming regions in the nearby galaxy NGC~6946.  
From a radio spectral decomposition, we find that the 33~GHz flux densities are typically dominated by thermal (free-free) radiation.  
However, we also detect excess Ka-band emission for an outer-disk star-forming region relative to what is expected given existing radio, submillimeter, and infrared data.
Among the 10 targeted regions, measurable excess emission at 33~GHz is detected for half of them, but in only one region is the excess found to be statistically significant ($\approx7\sigma$).    
We interpret this as the first likely detection of so called `anomalous' dust emission outside of the Milky Way.  
We find that models explaining this feature as the result of dipole emission from rapidly rotating ultrasmall grains are able to reproduce the observations for reasonable interstellar medium conditions.   
%
While these results suggest that the use of Ka-band data as a measure of star formation activity in external galaxies may be complicated by the presence of anomalous dust, it is unclear how significant a factor this will be for globally integrated measurements as the excess emission accounts for $\la$10\% of the total Ka-band flux density from all 10 regions.  
\end{abstract}

\keywords{radio continuum: general -- dust -- galaxies: individual (NGC~6946)} 

\section{Introduction}
Accurate separation of Galactic foreground emission in cosmic microwave background (CMB) experiments remains a major challenge in observational cosmology.  
Besides the standard foreground components, free-free, synchrotron, and thermal dust emission, an unknown component has been found to dominate over these at microwave frequencies between $\sim10-90$~GHz, and is seemingly correlated with 100~$\micron$ thermal emission from interstellar dust \citep[e.g.][]{ak96,ao97,el97}.  

The first detection of this `anomalous' dust-correlated emission was attributed to free-free emission from shock-heated gas in 
supernova (SN) remnants \citep{el97}.  
However, \citet{dl98a} demonstrated that the observed microwave excess could not be due to free-free emission from hot gas, as this scenario would require an unrealistically high energy injection rate (i.e. $\ga$2 orders of magnitude larger than that from SNe), and suggested two alternative models; electric dipole rotational emission from ultrasmall  ($a \la 10^{-6}$~cm) grains \citep{dl98b}, and magnetic dipole emission from thermal fluctuations in the magnetization of interstellar dust grains \citep{dl99}.  

At present, the most widely accepted explanation for the anomalous dust emission is the spinning dust model in which rapidly rotating very small grains, having a non-zero electric dipole moment, produce the observed microwave emission \citep{dl98b}.  
This model was recently refined by \citet{ahd09} who implemented the Fokker-Planck equation to compute grain angular velocity distributions using updated grain optical properties and size distributions.  
To date, anomalous dust emission has not been directly detected outside of the Galaxy.  
In this Letter we present the first likely detection for an outer-disk star-forming region in the nearby galaxy NGC~6946.  

\section{Observations and Data Analysis}
We targeted 10 star-forming regions in the nearby galaxy NGC~6946 \citep[$d \approx 6.8$~Mpc;][]{ik00}, including its starbursting nucleus, chosen due to their wealth of existing and forthcoming mid- and far-infrared spectroscopic data collected as part of the {\it Spitzer} Infrared Nearby Galaxies Survey \citep[SINGS;][]{rck03} and the project Key Insights on Nearby Galaxies: a Far-Infrared Survey with {\it Herschel} (KINGFISH; PI. R. Kennicutt).  
The location and identification number of each region is overlaid on the stellar continuum subtracted 8~$\micron$ image, which traces emission from very small grains (VSGs) and polycyclic aromatic hydrocarbons (PAHs), in the top panel of Figure \ref{fig-1}.  

\subsection{Radio Data}
\begin{figure}
\begin{center}
\scalebox{1.}{
\plotone{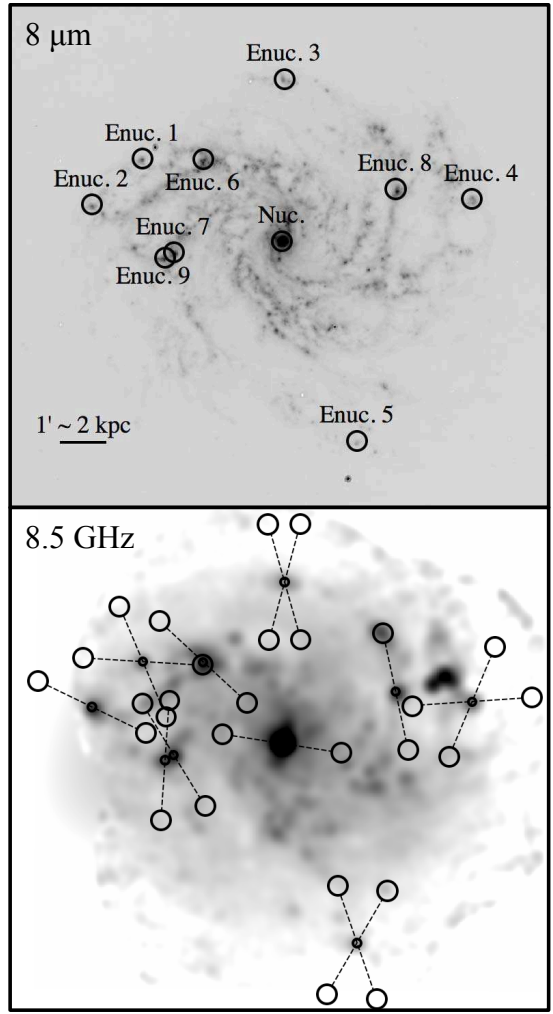}}
\caption{{\it Top}: 
The stellar continuum subtracted 8~$\micron$ map of NGC~6946 with the ID's of each targeted star-forming region.  
{\it Bottom}:  
The 8.5~GHz (15\arcsec beam) image from \citet{beck07} showing the location of the reference beam positions for each observation.  
\label{fig-1}}
\end{center}
\end{figure}

Observations in the Ka-band ($26 - 40$~GHz) were taken on 2009 March 21 using the Caltech Continuum Backend (CCB) on the 100~m Robert C. Byrd Green Bank Telescope (GBT).  
The CCB simultaneously measures the entire Ka bandwidth over 4 equally spaced frequency channels and synchronously reads out and demodulates the beamswitched signal to remove atmospheric fluctuation and/or gain variations.  
Reference beams are measured by nodding 1\farcm3 away from the source, and their positions are identified on the 8.5~GHz radio map in the bottom panel of Figure \ref{fig-1}.   
The average FWHM of the GBT beam in the Ka-band was $\approx$25\arcsec among our sets of observations, and is indicated by the size of each circle in Figure \ref{fig-1}; 
at the distance of NGC~6946 this projects to a physical scale of $\approx0.8$~kpc.      

A detailed description on the performance of the CCB receiver, along with information on the reduction pipeline and error estimates, can be found in \citet{bm09}.  
In addition to any systematic errors, a calibration error of 10\% was assigned to the flux density measurement at each frequency channel \citep{bm09}.    
To analyze the full radio spectral energy distributions (SEDs) of these regions, we compiled multifrequency radio data from the literature.  
The 1.4~GHz radio map ($14\arcsec \times12\farcs5$ beam) comes from the Westerbork Synthesis Radio Telescope (WSRT)-SINGS survey \citep{rb07} while the 1.5, 1.7, 4.9, and 8.5~GHz data ($15\arcsec \times 15\arcsec$ beam) all come from \citet{beck07}.  
The 4.9 and 8.5~GHz radio data included single-dish measurements.  

\subsection{Infrared and Radio Photometry}
All infrared data used here were included in the SINGS fifth data release.  
We make use of 8, 24, and 70~$\micron$ {\it Spitzer} data, as well as 450 and 850~$\micron$ SCUBA data \citep[see][]{dd05}, to compare the infrared and radio properties of each region as these data are at higher spatial resolution than the GBT data.  
Regions 2, 3, and 5 were not covered by the SCUBA 450 or 850~$\micron$ maps, while regions 4 and 8 were near the edges of the maps.    
The 8~$\micron$ map was corrected for stellar light using a scaled 3.6~$\micron$ image following \citet{gxh04}.  

Photometry was carried out on all radio and infrared maps after first cropping each image to a common field-of-view and regridding to a common pixel scale.  
To accurately match the photometry of our images to the GBT measurements, maps were convolved to the resolution of the GBT beam in the Ka-band.  
The {\it Spitzer} data were first resolution-matched to the 70~$\micron$ beam using custom kernels provided by the SINGS team.  
Since the {\it Spitzer} PSFs suffer from significant power in their side lobes, we convert to Gaussian PSFs using a standard CLEAN algorithm; the {\it Spitzer} PSF was removed and the image was restored with a Gaussian having a FWHM of 25\arcsec.  
Flux densities at each wavelength were then measured by taking the surface brightness at the location of each GBT pointing and multiplying by the effective area of the beam.  
The results from our radio photometry, along with 1~$\sigma$ error bars, are plotted in Figure \ref{fig-2} for all regions except extranuclear region 4, which are shown in Figure \ref{fig-3}.  

\subsection{Additional Considerations for the GBT Data}
Given that the beam size varies over the full Ka-band, we applied a correction to the flux densities at each frequency channel as if their beam were 25$\arcsec$.  
Scaling factors were derived by computing the photometry on the 8.5~GHz map using the beam sizes for each frequency channel averaged over the entire run.  
These correction factors are 0.89, 0.95, 1.00, and 1.04 at 27.75, 31.25, 34.75, and 38.25~GHz, respectively.  
The averaged Ka-band 
flux density, weighted by the errors from each channel over the full band, are given in Table 1 along with uncertainties; the corresponding effective frequency is $\approx$33~GHz  for each position (i.e. $32.83\pm0.39$~GHz, on average).        
Flux densities of the individual channels are plotted with 1~$\sigma$ error bars in Figures \ref{fig-2} and \ref{fig-3}.   

While the unblocked aperture of the GBT mitigates the significance of side lobes, 
elevation scans of the calibration source showed side lobes with an amplitude of $\approx$3\% of the beam peak.  
The added power may cause an overestimation of the 33~GHz flux density.  
Using a model beam having side lobes 
at 5\% of the beam peak, we recalculated the photometry for each radio, submillimeter, and infrared image finding that our results are not significantly affected by the 
presence of side lobes (see $\S$3).  

Since the reference beam throw is only 1\farcm3, real signal in our reference positions is a concern;  
having reference positions on the galaxy may result in an underestimation of the 33~GHz flux density (see bottom panel of Figure \ref{fig-1}).  
To assess the severity of this, we add back to the 33~GHz flux densities the average flux densities measured among the reference positions using two modeled 33~GHz maps by: 1) scaling the 24~$\micron$ data \citep[Equation 10 in ][]{ejm08}, and 2) scaling the 8.5~GHz data with the measured spectral indices from fitting the radio data shortward of 10~GHz.  
The minimum flux density of the reference beam positions for each target was then taken as the local sky and subtracted.  
Taking the median value among all reference positions appearing clearly off the disk as a global sky measurement yields very similar results.   
These `boosting' factors, taken as the weighted average between the values obtained using the two different modeled 33~GHz maps, are listed in Table \ref{tbl-1} and are typically small ($\la$10\%) with the exception of extranuclear regions 1, 5, and 8, for which one of the reference beams appeared to fall on an adjacent bright region.  
While the 8.5~GHz-derived 33~GHz flux densities at the reference positions are typically twice as large as the 24~$\micron$-derived values, which only estimate the free-free contribution, the corresponding `boosting' factors agree within 5\%, on average (i.e. $1.04\pm0.05$).  

\begin{deluxetable}{cccccc}
\tablecaption{Positions, K{\rm a}-Band Flux Densities, and Excesses \label{tbl-1}}
\tablewidth{0pt}
\tablehead{
\colhead{}  & \colhead{R.A.} & \colhead{Decl.} &
\colhead{$S_{\rm 33~GHz}$} & 
\colhead{$f_{\rm 33~GHz}^{\rm exc}$} & \colhead{}\\ 
\colhead{ID} & \colhead{(J2000)} & \colhead{(J2000)} &
\colhead{(mJy)} &\colhead{(\%)} &\colhead{corr$^{\dagger}$}
}
\startdata
     Nucleus   &20~34~52.34   &+60~ 9~14.2  &15.7$\pm$0.79  & -26$\pm$16&  1.00\\
     Enuc. 1   &20~35~16.65   &+60~11~ 1.1  & 0.4$\pm$0.05  &-137$\pm$47&  2.57\\
     Enuc. 2   &20~35~25.49   &+60~10~ 1.8  & 2.4$\pm$0.16  &  10$\pm$11&  1.05\\
     Enuc. 3   &20~34~51.89   &+60~12~44.8  & 1.0$\pm$0.07  &  14$\pm$15&  1.08\\
     Enuc. 4   &20~34~19.17   &+60~10~ 8.7  & 2.9$\pm$0.15  &  45$\pm$ 6&  1.06\\
     Enuc. 5   &20~34~39.27   &+60~ 4~55.1  & 0.5$\pm$0.05  & -22$\pm$28&  1.19\\
     Enuc. 6   &20~35~ 6.09   &+60~11~ 0.6  & 2.8$\pm$0.18  &  -9$\pm$13&  1.09\\
     Enuc. 7   &20~35~11.21   &+60~ 8~59.7  & 3.1$\pm$0.20  &  10$\pm$10&  1.03\\
     Enuc. 8   &20~34~32.52   &+60~10~22.0  & 1.8$\pm$0.14  & -18$\pm$15&  1.26\\
     Enuc. 9   &20~35~12.71   &+60~ 8~52.8  & 2.4$\pm$0.16  &   8$\pm$11&  1.05
\enddata
\tablecomments{$^{\dagger}$ Estimated $S_{\rm 33~GHz}$ correction factor due to over subtraction by reference beams.  These corrections were not included in the calculation of  $f_{\rm 33~GHz}^{\rm exc}$.}
\end{deluxetable}

\subsection{Modeling the Infrared-to-Radio Spectra \label{sec-modspec}}  
\begin{figure*}
\begin{center}
\scalebox{1.}{
 \plotone{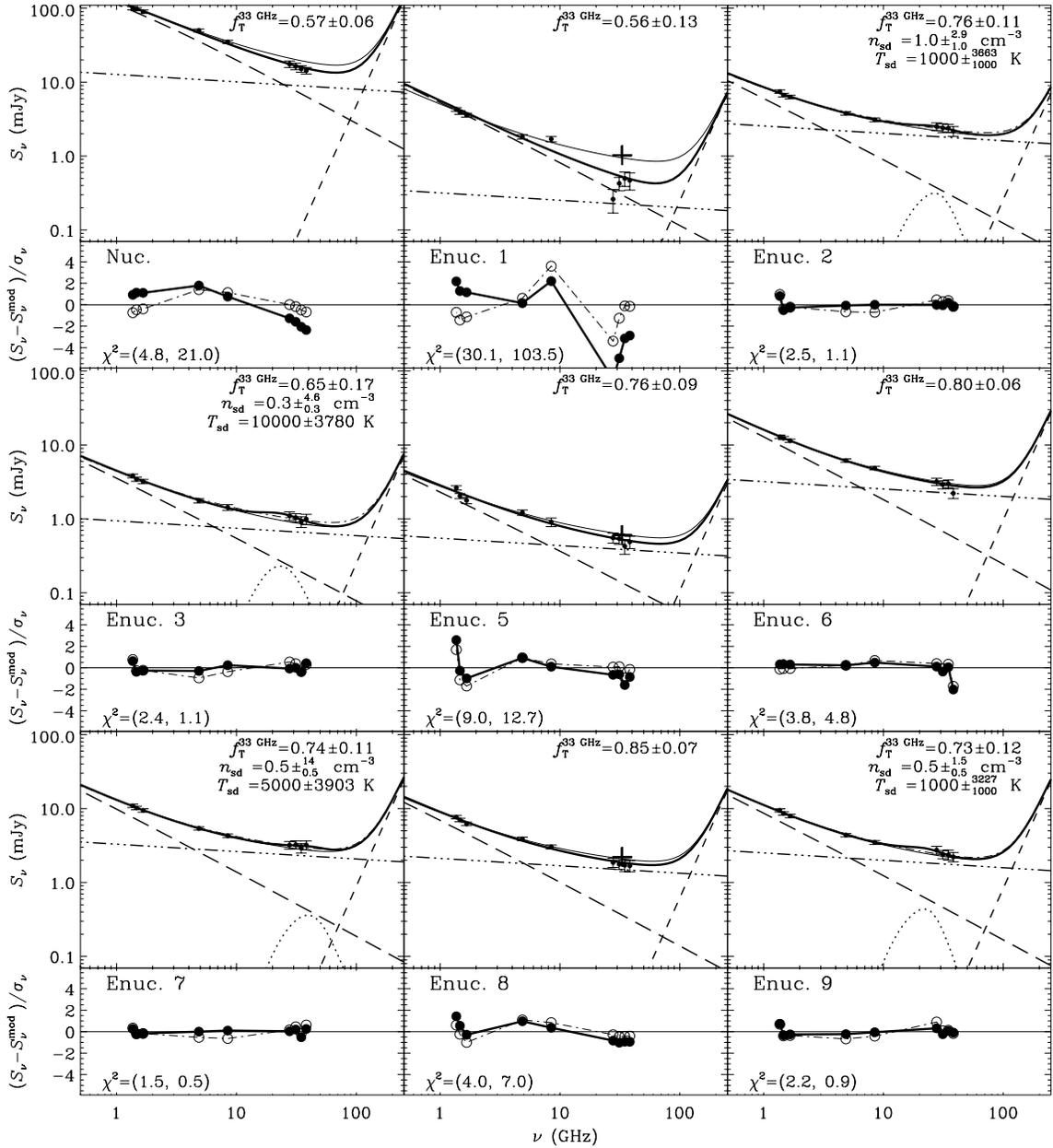}}
\caption{
The radio spectrum of each star-forming region excluding extranuclear region 4 (see Figure \ref{fig-3}).  
The best-fit model spectra 
are given by {\it thick solid} lines.  
The \citet{dh02} templates which best fit the infrared and submillimeter data are shown using the {\it short dashed}-lines.  
The free-free, non-thermal, and spinning dust components of the radio spectra are shown using {\it triple dot dashed}-, {\it long dashed}-, and {\it dotted}-lines, respectively.  
The best-fit models neglecting the anomalous dust component which fit all the radio data, and those data below 10~GHz, are given by the {\it dot dashed}- and {\it thin solid}-lines, respectively (See $\S$\ref{sec-modspec}).    
For the nucleus and extranuclear regions 1, 5, 6, and 8, the best-fit models overlap the {\it dot dashed}-lines since excess 33~GHz emission was not detected.
In the top right corner of each panel the 33~GHz thermal fraction from the best-fits are given 
along with the best-fit ISM densities and kinetic gas temperature for the spinning dust component under WNM or WIM conditions \citep{dl98b} where appropriate.  
For extranuclear regions 1, 5, and 8, where we believe a considerable amount of signal was subtracted out by the reference beams, a cross is placed at the position of the corrected flux density.  
Below each spectrum the signal-to-noise of the residuals are shown for the two- ({\it dot-dashed}-lines) and three-component ({\it thick-solid}-lines) models along with corresponding $\chi^{2}$ values (see $\S$4.1).  
\label{fig-2}}
\end{center}
\end{figure*}

\begin{figure}
\begin{center}
\scalebox{1.}{
\plotone{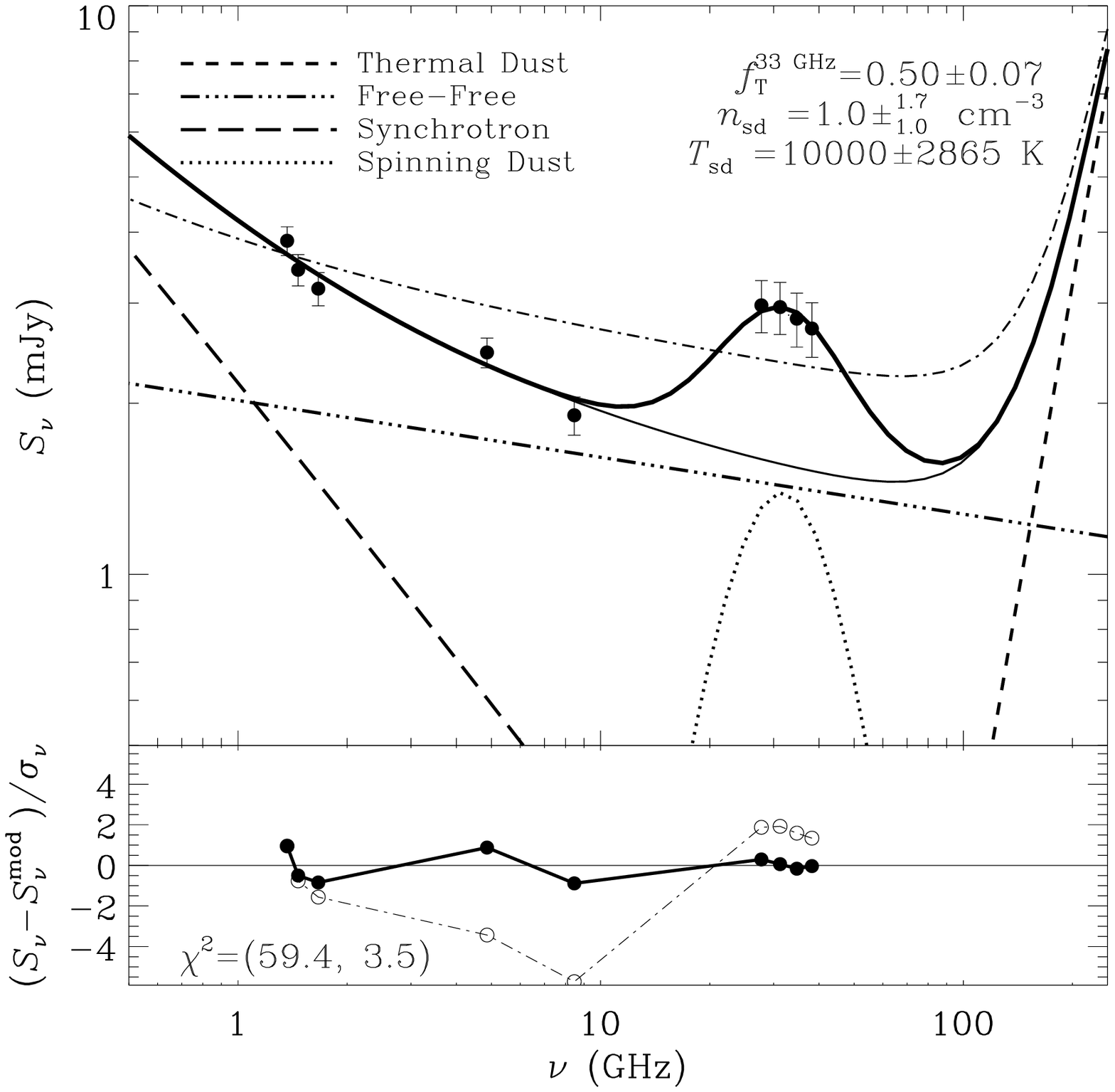}}
\caption{The radio spectrum of extranuclear region 4 along with signal-to-noise of the residuals for the best-fit models (see caption of  Figure \ref{fig-2}).  
\label{fig-3}}
\end{center}
\end{figure}

We fit the infrared and radio data independently using three separate components as described in \citet{ejm09}; thermal dust emission, thermal (free-free) radio emission, and non-thermal synchrotron emission (see Figures \ref{fig-2} and \ref{fig-3}).  
The infrared, and submillimeter data where available, were fit using the SED templates of \citet{dh02}.  
We then fit the radio data using a combination of thermal and non-thermal emission.  
Since we are interested in measuring excess emission in the microwave range, we perform two separate fits using: 1) all ($1.4-40$~GHz) radio data, and 2) only radio points at frequencies below 10~GHz since these frequencies are likely free of significant anomalous dust emission.  

We first assume a thermal (free-free) component which scales as $S_{\nu}^{\rm T} \propto \nu^{-\alpha_{\rm T}}$, where $\alpha_{\rm T}=0.1$ is the thermal spectral index.  
The non-thermal component is proportional to the energy loss of CR electrons to synchrotron emission relative to the total energy loss, scaling as $S_{\nu}^{\rm NT} \propto |dE/dt|_{\rm syn} N(E) \propto \nu^{-\alpha_{\rm NT}}$, where $N(E)$ is the number density of CR electrons per unit energy and $\alpha_{\rm NT}$ is the non-thermal spectral index.  
We assume an energy injection spectrum of $Q(E) = \kappa E^{-p}$, where $p=2.8$ for shock-accelerated electrons that have interacted with the ambient medium \citep[e.g.][]{bs93,jb09}.  
Our models of the non-thermal emission formally include energy losses to CR electrons arising from synchrotron, inverse Compton (IC) scattering, ionization, bremsstrahlung, as well as a prescription for escape; losses from advection are ignored \citep[see~][~for details]{ejm09}.    
For simplicity, we assume interstellar medium (ISM) densities of $n_{\rm ISM}=0.1$ and 10~cm$^{-3}$ for the extranuclear regions and nucleus, respectively.  
Consequently, ionization and bremsstrahlung losses are insignificant relative to synchrotron and IC losses for the extranuclear regions, and become comparable for the nucleus.  

Under these assumptions we generate the shape of the non-thermal radio spectrum, resulting in values of $\alpha_{\rm NT} \approx 0.8$, on average for the extranuclear regions, and $\alpha_{\rm NT} \approx0.7$ for the nuclear region,   
and vary the fraction of thermal emission by scaling the relative amplitude of the thermal and non-thermal components to best fit all radio points, and only those below 10~GHz, using a standard $\chi^{2}$ minimization procedure ({\it dot dashed}- and {\it thin solid}-lines in Figures \ref{fig-2} and \ref{fig-3}, respectively).
The corresponding best-fit model 33~GHz thermal fractions,  ranging between $\approx50-85\%$, are given in the top right corner of each panel along with their uncertainties.  
For reference, the global thermal fraction at $\approx$ 33~GHz is roughly 50\% among star-forming galaxies \citep{cy90}.  
Errors in the fitting and associated parameters are calculated by a standard Monte Carlo approach using the photometric uncertainties of the input flux densities.  

\section{Results: Excess Emission at 33~GHz}
We find that the Ka-band continuum is typically dominated by free-free emission, having a median  thermal fraction of $\approx$70\%. 
We also estimate whether there is 33~GHz emission in excess to what our fits to the lower frequency ($\nu < 10$~GHz) radio data would predict (i.e. \(f_{\rm 33~GHz}^{\rm exc} = (S_{\rm 33~GHz}^{\rm obs}-S_{\rm 33~GHz}^{\rm mod})/S_{\rm 33~GHz}^{\rm obs}\)) and attach an uncertainty (see Table 1).  
Half of the sample have positive excesses ranging between $\approx10-50\%$ (i.e. extranuclear regions 2, 3, 4, 7, and 9), but only one (extranuclear region 4) has a statistically significant excess ($\approx 7\sigma$).  
From our fitting, we find that $\approx$50\% of the observed flux density from this region arises from free-free emission while the other $\approx$50\% appears to be anomalous, consistent with \citet{cd09} who report comparable fractions of free-free and anomalous dust emission for two components of a Galactic H{\sc ii} region.  
Even when using the photometry obtained with the exaggerated side-lobes (see $\S$2.3), the excess in extranuclear region 4 is still detected at the $\approx 5\sigma$ level indicating the detection is robust.      
We also apply Peirce's criterion for outliers \citep{bp52} to the distribution of all excesses, corrected for the over subtraction of sky ($\S$2.3), finding extranuclear region 4 to be an outlier at a significance of $\approx$5$\sigma_{\rm pop}$.
%

For extranuclear region 1 we find that the measured flux density is significantly lower (i.e. $\approx$40\%) than predicted;  
we attribute this discrepancy to over subtraction of sky as one of the reference beams fell on extranuclear region 6.  
The corrected flux density, given by a cross in Figure \ref{fig-2}, is very close to the expected value.   

The choice of $p$ for the electron injection spectrum affects the average 33~GHz thermal fraction from the fits, along with the significance of the excesses.  
For example, choosing $p\approx2.2$, which is in the injection limit ($2.0 \la p \la 2.4$) for electrons freshly shock-accelerated by SN remnants, yields an average thermal fraction at 33~GHz of $\la$10\% and results in the excesses from regions 2, 3, 7 and 9 also being significant at $\ga$2.5$\sigma$ level.  
However, the $\chi^{2}$ values from these fits are $\ga$3 times larger, on average, suggesting that an injection-limit value for $p$ is inappropriate.   
On the other hand, setting $p\approx2.6$ gives similar $\chi^{2}$ values and a median 33~GHz thermal fraction of $\approx$60\%.  

\section{Discussion: The Nature of the 33~GHz Excess}
We report on the likely detection of excess microwave emission from an extranuclear star-forming region in NGC~6946.    
A rising spectrum at these frequencies can be produced by optically thick free-free emission observed from compact H{\sc ii} regions.    
However, this explanation appears inconsistent given the constraints set on the production rate of Lyman continuum photons ($N_{\rm Lyc}$) by the 24~$\micron$ data.  
Assuming that all of the 33~GHz flux density from extranuclear region 4 arises from free-free emission suggests $N_{\rm Lyc}\sim 1.4\times10^{52}~{\rm s}^{-1}$,   
a factor of more than 5 times larger than the $N_{\rm Lyc}\sim 2.6\times10^{51}~{\rm s}^{-1}$ rate inferred from the 24~$\micron$ data.  
Thus, explaining the excess 33~GHz emission by compact H{\sc ii} regions seems unlikely.  


It is at precisely these frequencies where excess emission, coined `anomalous' dust emission, has been found in diffuse emission experiments \citep[e.g.][]{rd06,sh07,df08}.   
Taking extranuclear region 4 as an example, we find that its 33~GHz flux density is better fit by including a component besides free-free, synchrotron and thermal dust emission,    
and its shape appears reasonably well constrained (see Figure \ref{fig-3}).  
On the high frequency side, the spectrum is constrained by measurements from the infrared into the submillimeter, and is most unlikely to exceed the \citet{dh02} model curves above 300~GHz.  
On the low frequency side, data from extranuclear region 4 requires the excess emission component to fall off from 30 to 10~GHz.  
This supports the spinning dust interpretation, which predicts a peaked component in this frequency range.  
We therefore believe that our observations indicate that an anomalous dust component exists, it is detectable in extragalactic sources, and it can be reproduced by spinning dust.        
To our knowledge, this is the first detection of anomalous dust emission outside the Galaxy.   

\subsection{The Spinning Dust Model}
Currently, the favored origin of this excess emission is electric dipole radiation from rapidly rotating ultrasmall grains \citep{dl98b}.  
Using the revised model of \citet{ahd09}, we fit all occurrences of excess Ka-band emission (see $\S$3) assuming a range of warm neutral medium (WNM; $x_{\rm H} \approx 0.01$) and warm ionized medium (WIM; $x_{\rm H}\approx 0.99$) conditions as defined by \citet{dl98b}, where $x_{\rm H}$ is the ionization fraction.  
The parameters that were allowed to vary were ISM density ($n_{\rm ISM} = 0.1,0.3,0.5,1,5,10,15,30$~cm$^{-3}$) and gas kinetic temperature ($T = 0.1,0.3,0.5,0.7,1\times10^{4}$~K).  
The spinning dust spectra were first scaled to match the observed excess 33~GHz flux density and then added to the best-fit thermal$+$non-thermal radio spectra from fitting the radio data below 10~GHz.  
The residuals of this three component model against the data, fitting each of the 4 CCB frequency channels independently, were minimized to pick the favored spinning dust component by varying the spinning dust parameters only.  
The signal-to-noise of the residuals, along with the corresponding $\chi^{2}$ values, are shown beneath each radio spectrum in Figure \ref{fig-2} for the cases of fitting all radio data with two components (i.e. thermal $+$ non-thermal radio emission) 
and this three component model.  

The emissivity of the best-fit spinning dust model for extranuclear region 4 corresponds to a hydrogen column density of ${\rm N_{H}} \approx 3.3\times10^{21}~{\rm cm^{-2}}$.  
This is a factor of $\approx$1.7 times larger than the  ${\rm N_{H}} \approx 1.9\times10^{21}~{\rm cm^{-2}}$ average column measured using H{\sc i} data from \citet{fw08}, and therefore  
consistent with the observed levels for a reasonable molecular gas fraction.   


While the excess at 33~GHz is only statistically significant for extranuclear region 4, the fits are improved by including the spinning dust component in extranuclear regions 2, 3, 7, and 9 as evidenced by the lower $\chi^{2}$ values. 
The ISM density and gas temperature from the best fitting spinning dust component are given and point to WNM parameters for each region.  
However, the lack of a fine spectral sampling, and large uncertainty in the Ka-band spectral index, yields a large range of acceptable values.    
It should also be noted that each of these regions are located near the disk-edge of NGC~6946.

\subsection{Implications}
These results suggest that the Ka-band observations from external galaxies may contain appreciable amounts of anomalous dust emission and that the assumption that this frequency range is dominated by free-free emission, making it a clean measure for the star formation in external galaxies, needs further investigation.  
It is currently unclear how much of an effect this component might have for interpreting globally integrated 33~GHz flux density measurements of galaxies.  
Summing the 33~GHz flux densities for all 10 regions, we find that the excess emission accounts for $\la$10\% of the total 33~GHz flux density even when correcting for badly placed reference positions.  
Future Ka-band observations at higher resolution using the EVLA should help to address this by precisely identifying the origin of the excess microwave emission (e.g. in the PDRs surrounding the H{\sc ii} regions), and in turn also allow for signifiant progress in modeling this component out of the Galactic foreground emission for CMB experiments.    

\acknowledgements
We thank the anonymous referee for their useful suggestions that helped to improve the content and presentation of this paper. 
EJM thanks William Reach and Roberta Paladini for stimulating discussions.  
The National Radio Astronomy Observatory is a facility of the National Science Foundation operated under cooperative agreement by Associated Universities, Inc.
We are grateful to the SINGS team for producing high quality data sets used in this study. 
Support for this work was provided by NASA through an award issued by JPL/Caltech.

\end{document}